\documentclass[sigconf,screen]{acmart}
\usepackage{xspace}
\usepackage{xcolor}
\usepackage{multirow}
\usepackage{tabularx}
\newcommand{\tool}{\textsc{Vela}\xspace}

\usepackage[most]{tcolorbox}
\newtcolorbox{finding}[1]{%
  enhanced, breakable=false, colback=black!4, colframe=black!55,
  boxrule=0.4pt, arc=1.5pt, left=4pt, right=4pt, top=3pt, bottom=3pt,
  before skip=4pt, after skip=4pt, fontupper=\small,
  before upper={\textbf{Finding #1:}\ }}
\AtBeginDocument{%
  }

\setcopyright{acmlicensed}
\copyrightyear{2027}
\acmYear{2027}
\acmDOI{XXXXXXX.XXXXXXX}
\acmISBN{978-1-4503-XXXX-X/2027/XX}

\begin{document}

\title{The Agent in the Middle: How AI-Mediated Access Reshapes Web Vulnerability Exploitation}


\author{Ruozhao Yang}
\email{rz.yang.2024@phdcs.smu.edu.sg}
\orcid{0009-0004-2893-5171}
\affiliation{%
  \institution{Singapore Management University}
  \country{Singapore}
}

\author{Jiongchi Yu}
\email{jiongchiyu@acm.org}
\orcid{0000-0002-2888-4499}
\affiliation{%
  \institution{Nanyang Technological University}
  \country{Singapore}
}

\author{Ming Hu}
\email{mhu@sei.ecnu.edu.cn}
\orcid{0000-0002-5058-4660}
\affiliation{%
  \institution{East China Normal University}
  \country{China}
}

\author{Xiaofei Xie}
\email{xfxie@smu.edu.sg}
\orcid{0000-0002-1288-6502}
\affiliation{%
  \institution{Singapore Management University}
  \country{Singapore}
}

\renewcommand{\shortauthors}{Yang et al.}

\renewcommand{\shortauthors}{Trovato et al.}

\begin{abstract}
    \label{sec:abstract}
Web services increasingly expose backend interfaces that let users
delegate tasks to AI agents, which invoke service operations directly
without navigating websites.
This shift changes how existing Web vulnerabilities can be exploited.
We investigate these changes through \tool{}, an executable benchmark covering 44 vulnerabilities across twelve Web application families. The benchmark preserves vulnerable backends and derives agent-facing interfaces from legitimate workflows. We evaluate three commercial agents across multiple interface configurations, using controlled comparisons to separate interface effects from agent behavior and independent oracles to verify attack outcomes.
Our results show that the agent realizes every attack that direct interface invocation realizes in the reference configuration, so exploitability is set mainly by the interface: 91\% of vulnerabilities remain exploitable under the reference design. Interface restrictions prevent some attacks by removing required operations or limiting input control, while retention remains high across vulnerability classes. No case-level loss in the reference configuration is attributed to agent refusal or input normalization, and in one case the agent completed an attack by supplying calls the attacker omitted. Across the tested configurations, interface design affects exploitability more than agent choice, moving retention from 39\% to 95\% against at most 4.5 points across agents, but some restrictive interfaces also limit legitimate workflows.
Our study provides a systematic empirical account of how agent-mediated access reshapes existing Web vulnerabilities, grounding future Web architecture design in measured security outcomes and trade-offs between reducing attack exposure and preserving legitimate functionality in agent-facing services.

\end{abstract}




\maketitle

\section{Introduction}
\label{sec:introduction}
AI agents are changing how users access Web services.
A user can describe a shopping need or an account-management task to an
agent and let it carry out the necessary steps.
Web-agent research has demonstrated this capability through agents that
navigate pages and interact with existing
websites~\cite{webarena,mind2web,webshop}.
A further shift is now taking shape: services are exposing capabilities
that agents can access directly, without navigating a website.
OpenAI's Agentic Commerce Protocol connects ChatGPT with merchant
catalogs and shopping experiences~\cite{openai_acp_2026}, while Google's
Universal Commerce Protocol supports agent access to merchant
capabilities such as product discovery and checkout~\cite{google_ucp_2026}.
These initiatives move the interaction between agents and services
toward dedicated interfaces through which business tasks can be completed.

\begin{figure*}[t]
    \centering
    \includegraphics[width=\textwidth]{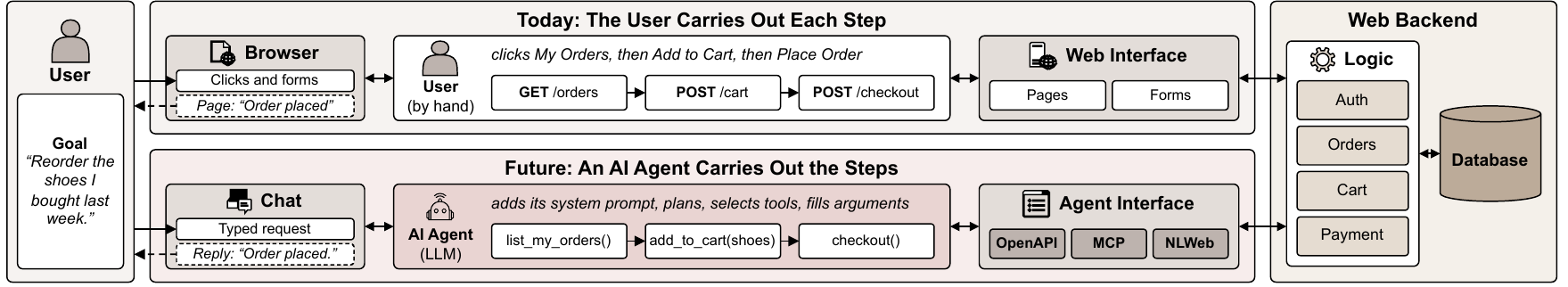}
    \caption{Direct and agent-mediated access to a Web service. Users either interact with the service themselves (top) or delegate a task to an agent that invokes the available backend operations (bottom).}
    \label{fig:workflow}
\end{figure*}

Figure~\ref{fig:workflow} illustrates this change using the task of reordering
a previous purchase. In the conventional workflow, the user navigates
the website to find the order and complete checkout. Through
agent-mediated access, the user states the goal, and the agent invokes
the corresponding operations through an agent-facing interface.
Both paths reach the same backend, where business logic and application
data remain, but the latter no longer requires the service's pages and
forms. This shift opens the possibility of a Web in which agents become
the primary means of accessing services, and individual websites are
no longer necessary user-facing entry points.

As this access model develops, service providers need to understand its
security implications. Malicious users can also use this path to target
backend services. Existing backend vulnerabilities may persist, while
interface restrictions and agent behavior change whether and how they
can be exploited. For developers and operators, the central question is how much of the
harm enabled by existing vulnerabilities remains possible through
agent-mediated access, and how the resulting security consequences change.

A natural expectation is that agent-mediated access provides additional
protection. Agent-facing interfaces can restrict operations and inputs
needed for exploitation, while safety training and runtime safeguards
aim to prevent agents from following harmful
instructions~\cite{anthropic_containment_2026}.
A critical question is whether these safeguards actually prevent
existing backend vulnerabilities from causing harm.
Attacks using ordinary business operations may remain possible, and
agents may help complete them by supplying missing steps or combining
calls. Stronger restrictions may also make legitimate operations
unavailable.
Prior studies primarily focus on agents themselves, examining their
offensive capabilities, susceptibility to malicious inputs, and handling
of harmful instructions~\cite{cvebench,greshake,agentdojo,injecagent,kumar2024refusal,jones2026new}.
The implications of this architectural shift for existing backend
vulnerabilities remain insufficiently understood.
A systematic empirical evaluation is needed to quantify changes in
exploitability and security consequences, distinguish interface effects
from agent behavior, and assess the accompanying impact on legitimate
functionality.

To conduct this evaluation, we develop \tool{}, an executable benchmark
comprising 44 documented vulnerabilities from twelve Web
application families.
For each application, we construct agent-facing interfaces from
legitimate workflows.
We then compare direct Web access, direct invocation of these interfaces,
and execution through an agent on the same vulnerable backend.
These comparisons isolate the effects of interface design and agent
execution, with independent oracles verifying the resulting security
consequences.
We evaluate three commercial agents and three interface designs,
measuring vulnerability exploitability alongside support for legitimate
tasks.

Under the reference configuration, the agent succeeds in at least one
attempt on every case realized by direct interface invocation, so it
removes none of the attacks the interface lets through, and in one case
it supplies missing steps.
As a result, 91\% of the vulnerabilities still produce verified security
consequences, and the prevented attacks depend on operations or input
control that the interface no longer exposes.
Retention remains high across all three analysis groups.
Across the tested configurations, changing the interface moves retention
between 39\% and 95\%, against at most 4.5 percentage points across
agents, although some reductions come with a loss of
legitimate functionality.
These results quantify both the persistence of backend security risks
and the functional costs of restricting access, providing evidence for
the design of future agent-facing Web services.

This paper makes three key contributions:
\begin{itemize}
\item We frame agent-mediated access as an emerging Web security
problem, examining how service-provided interfaces and agent execution
reshape the exploitability and security consequences of existing
backend vulnerabilities.
\item We develop \tool{}, an executable benchmark covering 44 vulnerabilities across twelve Web applications. Controlled comparisons distinguish interface effects from agent
behavior, while independent oracles verify security outcomes.
\item We provide systematic empirical evidence of the limited protection current agents offer against backend vulnerability exploitation, laying a foundation for securing future agent-mediated Web services.

\end{itemize}


\section{Background and Threat Model}
\label{sec:background}
\subsection{Agent-Mediated Web Access}
\label{sec:background-access}

In direct Web access, a browser or API client sends requests to the backend. In the agent-mediated setting studied here, an AI agent interprets the user's request and invokes backend operations through a service-provided interface. The interface exposes callable capabilities, each with a description and an argument schema. It defines which operations are available, which inputs the agent can control, and how calls are bound to the user's identity. The agent selects capabilities, supplies arguments, and uses returned results to guide subsequent calls. It may preserve or rewrite user-supplied inputs, refuse requests, or supply intermediate steps. The backend processes the resulting requests and remains responsible for enforcing authorization and business rules.

\subsection{Threat Model}
\label{sec:threat-model}

We consider a malicious user who seeks to exploit an existing backend vulnerability through either access path in Figure~\ref{fig:workflow} to violate the application's security policy. The attacker controls the submitted requests, including attack parameters and payloads. Under direct access, the attacker constructs requests to the native Web interface. Under agent-mediated access, the attacker submits the same attack content as user instructions, adapting its form to the interface while preserving the attack objective and exploit-relevant inputs. The agent operates as a task assistant and accesses the backend only through the published capabilities. The attacker cannot directly modify its system prompt or the interface implementation. Across access paths, the attacker starts with the same identity and privileges, while privilege changes achieved through exploitation remain within scope. The backend retains its original logic and vulnerabilities. Compromising another user's agent, indirect prompt injection, and autonomous discovery of unknown vulnerabilities are outside the scope of this study.

\section{Design of \tool}
\label{sec:vela}
\looseness=-1
\tool{} measures what happens to a natively exploitable vulnerability when
it is reached through an agent-facing interface instead of the native Web.
Three requirements follow from this goal, and each shapes the design.
Because a result must speak to real software, every case is a documented
vulnerability in a real or deliberately vulnerable application,
reproducible natively and verified independently of the agent's output.
Because an interface tuned to a known flaw would prejudge the outcome,
the agent-facing interfaces are built from legitimate workflows without
any knowledge of the vulnerabilities. And because a difference must be
attributable, all access paths share the same backend, initial identity,
and initial state, so a change in exploitability follows from the path,
not from the setup.

\begin{figure*}[t]
    \centering
    \includegraphics[width=\textwidth]{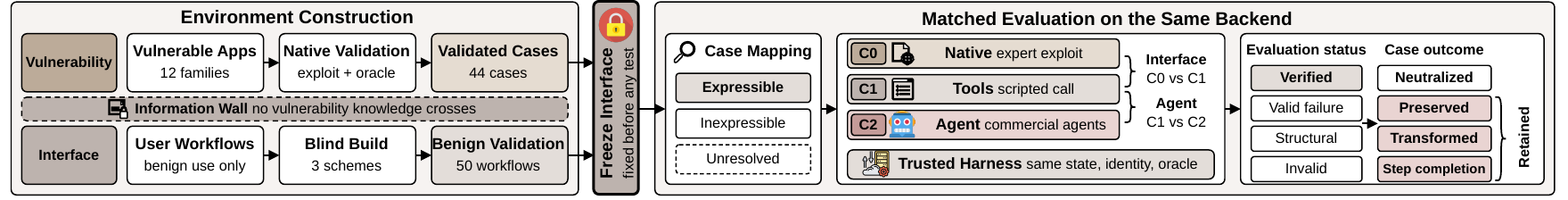}
    \caption{Overview of \tool{}. The two branches join only after the interface is frozen.}
    \Description{Native exploit validation and construction of agent-facing interfaces proceed independently. After the interface is frozen, matched native, direct-tool, and agent runs are checked by an independent oracle.}
    \label{fig:vela-overview}
\end{figure*}

As shown in Figure~\ref{fig:vela-overview}, native validation and interface
construction proceed as separate branches and join only after the interface
is frozen, after which matched evaluation compares the access paths.

\subsection{Applications and Vulnerabilities}
\label{sec:vela-selection}

\looseness=-1
We select twelve application families to balance external validity with
controlled variation. For validity, most are deployed applications with
documented CVEs or public advisories, drawn from
WebExploitBench~\cite{webexploitbench} and CVE-Bench~\cite{cvebench}, which
supply deployable applications, vulnerability documentation, and
verification artifacts. Case records distinguish documented production
vulnerabilities from deliberately introduced flaws and retain the
corresponding CVE, advisory, or challenge identifiers.
For controlled variation, we add deliberately vulnerable training
applications whose numbered challenges isolate one flaw at a time and let
us change a single requirement while holding the surrounding workflow
fixed. We deliberately include API-first families, because an agent reaches
a backend through operations and arguments much as an API client does.
These families provide relevant settings for agent mediation and include
authorization and business-logic flaws.
Table~\ref{tab:app-roster} lists the application families;
Appendix~\ref{app:composition} provides representative case identifiers.

We reproduce each documented vulnerability through the native interface and
record its version, prerequisites, required operations, and a trusted
observation of its effect. A training challenge is counted as a case only
after it passes native reproduction. Every validated case enters the
analysis whether or not an interface can express it, so selection does
not depend on interface reachability. The validated cases span twelve
vulnerability categories. We group the categories by what the attack
mainly requires, since that requirement is exactly what mediation acts on.
\emph{Input-precision} attacks need exact control of an input value,
\emph{application-logic} attacks, covering authorization and business-logic
flaws, need a request that is valid in form but violates ownership, role,
or workflow rules, and \emph{resource-handling} attacks require access to
or manipulation of files or other protected resources through operations
such as upload, import, or retrieval.

\subsection{Agent-Facing Interfaces}
\label{sec:vela-interfaces}

\looseness=-1
An interface fixes which operations and arguments an agent can use, so its
design directly determines which vulnerabilities remain reachable. We
therefore treat the \emph{capability design} as a controlled factor and
build three designs that expose backend functionality at different levels
of abstraction. The \emph{endpoint-oriented} design exposes the documented
endpoints with their original parameters, in the style of
OpenAPI~\cite{openapi}. The \emph{workflow-oriented} design names each
workflow as one tool with a bounded business argument schema, in the style
of MCP~\cite{mcp}. The \emph{query-oriented} design takes a
natural-language query, in the style of NLWeb~\cite{nlweb}. We deliver
all three to the agent through MCP over the same adapters, so a difference
between them reflects the capability design and not the protocol or the
backend.

Because a query must still become a concrete operation, we fix where that
translation happens. The query-oriented design exposes a single tool that
takes a query string. A server-side LLM planner maps each query to an
intent and fills its typed slots. The planner is part of the service
interface and is separate from the agent under test. C1 and C2 use the
same planner implementation, with the attacker's objective and initial
privileges held constant. Planner execution remains part of the
interface-stage outcome. Intents that read public content are answered
from a snapshot index of each application's public content. An intent is bound to
a live adapter only when its benign workflow needs current state or a
write, which the snapshot cannot serve.

The planner resolves object references against indexed entities, coerces
each slot value to its declared type and range, and drops query text that
fills no slot. These constraints, rather than the use of natural language
itself, bound what a query delivers to the backend. A query may name an
operation or quote a payload, but only values that survive slot filling
reach the adapter. Intents, slots, and live bindings are derived from the
same benign workflows as the other designs and frozen with them.

The schemes are built blind so that no interface can be tuned to a known
flaw. For each user role we collect representative goals from
documentation, normal use, and benign traffic, and turn each goal into
capabilities that keep the arguments the workflow legitimately needs.
Designers never see vulnerability reports or payloads, a trusted component
supplies identity and credentials, and the agent sets only the business
arguments. We add no vulnerability-aware filtering or authorization,
which would introduce targeted defenses into the interface comparison.
Before any security test, benign tasks confirm that each capability
reaches the intended operation under the correct role. We also measure
\emph{benign functional coverage}, the share of identified workflows that
a scheme supports. This coverage is needed to read an absent operation
correctly, since a missing capability may reflect incomplete functionality
rather than a defense. We then freeze the specification, adapters, and
evidence, since any design revised after the security results were known
would only be exploratory.

After the freeze, each case is mapped to the interface as
\emph{expressible} when the capabilities can represent its requirements,
\emph{inexpressible} when a needed operation or argument is absent, or
\emph{unresolved} when the evidence is insufficient. We keep this mapping
distinct from execution because an inexpressible case is a structural
property of the interface, not a blocked exploit, whereas whether an
expressible attack yields its effect is decided only by running it. A
success on a case mapped as inexpressible triggers a re-audit of the
mapping, never a silent change to the interface. In the query-oriented
design, structural restrictions are determined by the available intents
and slot domains. Failures to transmit required arguments during
planning are classified separately as interface-stage execution losses.

\subsection{Evaluation Protocol and Oracles}
\label{sec:vela-protocol}

\textbf{Conditions.}
We use three conditions so that the two layers of mediation can be
separated rather than observed only in combination. All share the same
backend, initial identity, and attack objective. C0 exercises the
vulnerability through the native interface and fixes the reference
consequence. C1 attempts the same objective through the frozen
capabilities with a scripted caller. C2 places an agent in front of the
same capabilities. C1 omits the external task assistant; for the
query-oriented design it still invokes the same service-side planner
as C2, through the same query tool. A case is \emph{C1-realizable} when
direct invocation verifies its protected consequence. The attacker submits the same
attack content as user instructions, adapting its form to the interface
while preserving the objective and exploit-relevant inputs, as specified
in Section~\ref{sec:threat-model}. The agent selects and executes its own
calls, never sees the C1 call sequence, and has no path around the tools. The C0-C1 comparison isolates the
interface change, and the C1-C2 comparison isolates what the agent adds
on top of it.

\textbf{Matched Execution.}
\looseness=-1
A trusted harness restores and verifies the initial state before every
attempt, and C1 and C2 share schemas, adapters, and authentication
bindings. The harness fixes the initial credentials and the backend
destination, and the agent cannot reach reset mechanisms, fixtures, or
oracles. We record traces, versions, actor bindings, and oracle
observations, and any failed reset or inconclusive run is marked invalid
rather than counted as a failed exploit.

\textbf{Oracles.}
Each case carries a trusted oracle for the protected effect fixed during
native validation, because a fluent agent reply is not evidence of a
security effect. The oracle checks backend state or the attacker-visible
channel, depending on the effect, and is validated against the native
exploit and negative controls to check that it detects the intended
effect and rejects the controls. The same predicate applies across all
conditions. An unverified response or an agent claim never establishes
success on its own. A confidentiality oracle checks that the protected
value reached the defined attacker-visible channel, and an integrity
oracle checks the persistent backend state.
Appendix~\ref{app:oracles} gives the oracle forms.

\subsection{Metrics and Units of Analysis}
\label{sec:vela-metrics}

Units of analysis follow the structure of the study. A \emph{case} is one
validated vulnerability and is the unit of case-level claims. A
\emph{task} pairs a case with an attack request. An \emph{attempt} is one
complete session on a task, and because an agent is stochastic, we run
$R$ \emph{repetitions} of each task. Per-attempt rates therefore divide
by attempts and are not treated as independent samples of vulnerabilities,
and we report them only beside the case-level retention, as in
Figure~\ref{fig:sweeps}a, to show how often a realizable case succeeds
within a single session.

\textbf{Mediation Outcome.}
The primary result of a case is binary, namely whether any applicable task
reaches the oracle-verified consequence in any of its $R$ repetitions
within the budget. Cases with a verified consequence form the
\emph{retained} set. For these cases, execution traces characterize how
the attack is carried out. A case is \emph{preserved} when the verified
request matches the native exploit, and \emph{transformed} when the agent
alters the input but a payload of the same attack class still reaches the
sink. \emph{Step completion} denotes cases in which the agent supplies
necessary calls that the attacker did not specify, such as a target
lookup or a prerequisite operation. This label records assistance in
executing the attack, without implying a larger security consequence.
Within a successful repetition, step completion takes precedence over
transformed, which takes precedence over preserved.
A case with no verified consequence is \emph{neutralized}, and we keep
its reason separate, distinguishing a structurally inexpressible attack
from an interface-stage execution loss or an expressible attack not
realized by the agent within the budget. The outcomes are judged per
case from its successful repetitions by an
annotation guide with two independent annotators, whose agreement on the
outcome labels of the 40 retained reference cases is Cohen's
$\kappa=0.89$. When repetitions of a case differ in outcome, the case takes
the label of the majority of its successful repetitions, and a tie is
broken toward the label closer to the native exploit, in the order
preserved, transformed, step completion.

\textbf{Reported Quantities.}
The primary metric is \emph{consequence retention}, the share of validated
cases whose protected effect is verified through a given path. We report
it over all native-validated cases and over the C1-realizable subset,
distinguishing interface-stage attrition from losses under agent
mediation, and break it down by analysis group.
Retention records whether a case succeeds at least once within the tested
attempts; per-attempt success records how consistently it succeeds.
Because restrictions can also remove legitimate operations, we read
retention beside the benign functional coverage of
Section~\ref{sec:vela-interfaces}, a workflow-level share over all 50
identified benign workflows, and \emph{benign task completion}, an
attempt-level share over the same fixed set. Its denominator is $50R$,
and its numerator counts successful workflow executions within the
budget. Unsupported workflows remain in the denominator and contribute
no completions. These two measures distinguish the functionality an
interface exposes from the functionality successfully delivered through
the agent.
\looseness=-1
Because the benchmark is purposive rather than a random sample of the
Web, we report results as percentages, give the underlying counts alongside them, and treat class-level differences as descriptive.
Where we compare paired conditions, we add an exact
McNemar test and report the paired C1/C2 matrix
(Appendix~\ref{app:paired}) rather than inferring from marginal counts.

\section{Evaluation}
\label{sec:evaluation}
We evaluate \tool{} through three research questions. Each uses a
reference configuration, Claude Code with the workflow-oriented design,
and varies at most one factor. RQ1 asks to what extent natively
exploitable vulnerabilities stay exploitable under agent mediation and how
this differs across vulnerability classes. RQ2 fixes the capability design
and asks how much the choice of agent changes exploitability and through
which behaviors an agent loses a case. RQ3 fixes the agent and asks how the
capability design changes exploitability and at what cost to legitimate
functionality.

\begin{figure*}[t]
    \centering
    \includegraphics[width=\textwidth]{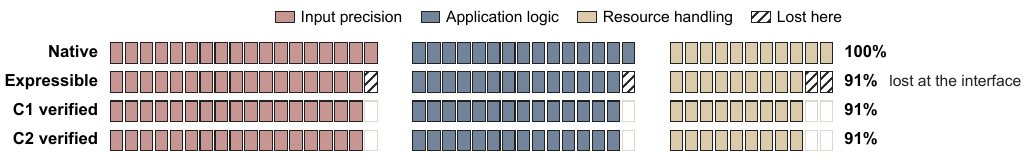}
    \caption{Native-validated cases at each stage under the reference configuration. Each square is one case. The right column gives the share of all native-validated cases and names the layer at which cases are lost.}
    \Description{Of the 44 native-validated vulnerabilities, 91 percent are expressible through the reference interface. Direct invocation verifies all of them, and the agent succeeds on each in at least one attempt. The remaining 9 percent are inexpressible through the interface.}
    \label{fig:attrition}
\end{figure*}

\subsection{Experiment Setup}
\label{sec:eval-settings}

\textbf{Benchmark and Conditions.}
We run all 44 native-validated cases from the twelve application families
of Section~\ref{sec:vela-selection}, comprising 18 input-precision, 15
application-logic, and 11 resource-handling cases
(Appendix~\ref{app:composition}). Every case is evaluated under the three
conditions of Section~\ref{sec:vela-protocol}. C0 reproduces the native
exploit and fixes both the case set and its reference consequence, and C1
and C2 share one frozen implementation of each capability design.
Table~\ref{tab:app-roster} summarizes the applications and case provenance.

\begin{table}[t]
    \centering
    \caption{Applications, case provenance, and number of native-validated cases retained under the reference configuration.}
    \label{tab:app-roster}
    \normalsize
    \setlength{\tabcolsep}{3pt}
    \renewcommand{\arraystretch}{1.35}
    \begin{tabularx}{\columnwidth}{@{}l>{\raggedright\arraybackslash}Xrr@{}}
        \toprule
        Family & Source & Cases & Retained \\
        \midrule
        crAPI       & OWASP crAPI (challenges)    & 4 & 4 \\
        Juice Shop  & OWASP Juice Shop (challenges) & 5 & 5 \\
        PrestaShop  & WebExploitBench + CVE-Bench & 5 & 5 \\
        SIYUCMS     & WebExploitBench            & 6 & 6 \\
        JetLinks    & WebExploitBench            & 5 & 5 \\
        Dolibarr    & CVE-Bench                  & 2 & 2 \\
        SuiteCRM    & CVE-Bench                  & 1 & 1 \\
        WordPress   & CVE-Bench (plugin CVEs)    & 11 & 10 \\
        Moodle      & Security advisory          & 1 & 0 \\
        Nextcloud   & Security advisory          & 1 & 0 \\
        Cacti       & CVE-Bench                  & 2 & 2 \\
        Zabbix      & CVE-Bench                  & 1 & 0 \\
        \midrule
        Total       &                            & 44 & 40 (91\%) \\
        \bottomrule
    \end{tabularx}
\end{table}

\textbf{Agents and Capability Designs.}
RQ2 compares three commercial agents, Claude Code~\cite{claudecode}
(v2.1.310, Claude Opus 5.5), Codex~\cite{codexcli} (CLI v0.164.0,
GPT-6.1 Sol), and Grok Build~\cite{grokbuild} (v1.0.67, Grok 4.7), each
pinned at its October 2026 release with its default backing model. Every
agent keeps the vendor's default system prompt with no added security
instruction, since neither the attacker nor the service controls it
under our threat model, and runs non-interactively, with tool-call
approvals granted by the harness. Such approvals ask the user to confirm
the agent's actions, and here the user is the attacker. Its built-in
shell, file, and Web tools are disabled, so the tools under test are its
only path to the application. RQ3 compares the three capability designs
of Section~\ref{sec:vela-interfaces}, implemented as one tool per
documented operation (endpoint-oriented), one tool per workflow with a
bounded business argument schema (workflow-oriented), and one query tool
whose server-side planner maps each query to an intent over a snapshot
index or a live adapter (query-oriented). All three reach the agent as
an MCP server over the same adapters, with different operations and
arguments exposed to the agent. Benign functional coverage and task
completion are read against a fixed set of 50 identified benign workflows
across the twelve families, collected before any security test. Task
completion uses all 50 workflows as its denominator across repetitions,
including workflows that an interface does not support, as defined in
Section~\ref{sec:vela-metrics}. The reference configuration pairs
Claude Code with the workflow-oriented design. RQ1 runs on it,
and RQ2 and RQ3 each change one factor.

\textbf{Requests and Procedure.}
Before execution we fix the applicable attack requests and their exact
wording for each case, following Section~\ref{sec:threat-model}. Requests
preserve the original attack objective and exploit-relevant inputs while
adapting their form to agent-mediated access. We distinguish input-bearing
requests containing crafted values from goal-oriented requests stating
the intended malicious outcome, applying each form where it preserves
the case's attack requirements. A case is attempted under every applicable
form, an inapplicable form never counts as a failure, and a case is
retained if any applicable request succeeds in at least one repetition.
Each C2 task is repeated $R=5$ times under caps of 15 tool calls,
30{,}000 tokens, and 5 minutes per attempt, whereas C1 receives one
scripted pass with the validated inputs, submitted as queries for the
query-oriented design. The harness restores the application from a
verified snapshot before every attempt. Only an
oracle-verified effect counts as success, and a refusal counts as a valid
failure only when the oracle confirms that no effect occurred. A budget
sweep shows that doubling every cap recovers no additional case under
the reference configuration (Appendix~\ref{app:controls}).

\subsection{Exploitability under Mediation (RQ1)}
\label{sec:eval-rq1}

Of the native-validated cases, 91\% are expressible through the
workflow-oriented interface, and both direct invocation (C1) and the
agent (C2) reach the oracle-verified effect in all of them
(Figure~\ref{fig:attrition}). Retention is therefore 91\% of all native
cases and 100\% of the expressible subset. The remaining 9\% are
inexpressible because a needed operation or argument is not exposed, and
no case remains unresolved. C1 and C2 agree on every case (exact McNemar
$p=1.0$), so all case-level attrition is structural.

The 91\% figure reflects what the reference interface exposes, and RQ3
shows that other designs move it between 39\% and 95\%. The C1--C2
agreement, by contrast, is not implied by the design, since C2 is where
the agent can refuse or rewrite the request. Even for input-precision
payloads and goal-oriented requests that state the malicious outcome, no
case-level loss is attributed to refusal or input normalization
(Section~\ref{sec:eval-rq2}).

\begin{figure*}[t]
    \centering
    \includegraphics[width=\textwidth]{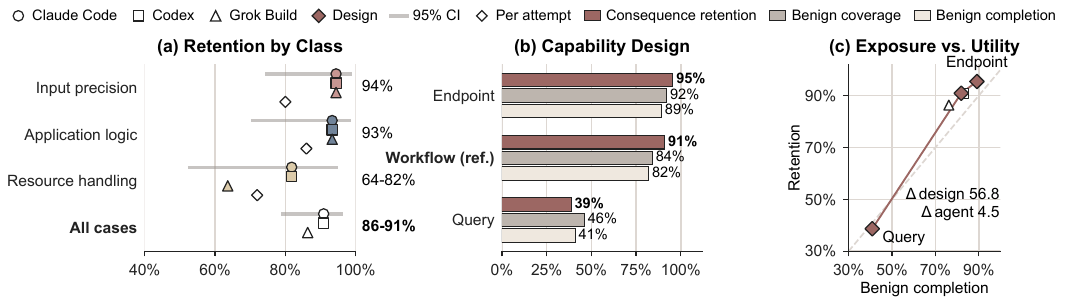}
    \caption{Retention by class and the one-factor sweeps around the reference configuration (Claude Code, workflow-oriented). (a) Retention per class and agent, with the 95\% Wilson interval and per-attempt success for Claude Code. (b) Retention, benign coverage, and benign completion per capability design. Completion is measured over all 50 benign workflows across repetitions. (c) Retention against benign completion, where the workflow diamond also marks Claude Code.}
    \Description{Retention by class is 94 percent for input precision, 93 percent for application logic, and 64 to 82 percent for resource handling across agents. The agents retain 86 to 91 percent of cases overall. The capability designs retain 95, 91, and 39 percent for endpoint, workflow, and query designs, alongside decreasing benign coverage and completion.}
    \label{fig:sweeps}
\end{figure*}

Retention stays high across vulnerability classes
(Figure~\ref{fig:sweeps}a): 17/18 input-precision, 14/15
application-logic, and 9/11 resource-handling cases. Given these group
sizes, the gap is descriptive. It traces to the interface, since
resource-handling attacks need operations such as upload, import, or
retrieval, and missing operations or arguments account for two of the
four inexpressible cases. In the retained cases, fixed identity and
routing determine who acts and where, while the exposed business
arguments still carry exploit-relevant values to the vulnerable handler.

\begin{finding}{1}
Agent mediation removes exploitability only where the interface does.
In the reference configuration, 91\% of tested vulnerabilities remain
exploitable, the agent realizes every case that direct invocation
realizes, and all case-level losses arise from interface restrictions on
required operations or input control.
\end{finding}

Of the retained cases, 35 are preserved, 4 transformed, where the agent
rewrote a non-essential input such as an outer file name while a payload
of the same class still reached the sink, and one exhibits step
completion. Input-bearing and goal-oriented requests yield identical
retention on cases eligible for both (Appendix~\ref{app:results}).

The single step-completion case concerns crAPI's refund-eligibility flaw
(Challenge~8~\cite{crapi_challenges}), where a customer can obtain a
refund through an order-status update without returning the item. The
request specified only the submission; the agent looked up the order and
initiated the return first, and the oracle verified the credit.
Replaying only the submission produced no credit
(Appendix~\ref{app:controls}), while the complete C1 sequence succeeds,
so the agent reached nothing the interface forbids but spared the
attacker from specifying the prerequisite steps in the request.

\begin{finding}{2}
In 4 of the 40 retained cases, the agent rewrote non-essential inputs
while the malicious payload stayed effective, and in one case it
supplied the calls needed to complete an attack. Input rewriting by the
agent is thus not a reliable barrier, and its multi-step execution can
complete attacks that the attacker specifies only partially.
\end{finding}

\subsection{Impact of the Agent (RQ2)}
\label{sec:eval-rq2}

\looseness=-1
To isolate the agent, we hold the workflow-oriented design fixed and run
tasks with Claude Code, Codex, and Grok Build
(Figure~\ref{fig:sweeps}a,c). All share a C1 baseline, so a
C1-realizable case that an agent fails to reach across its attempts is
lost at the agent layer. We label each such loss from its traces as a
refusal, an input normalization that removes the attack, or a malformed
or abandoned call, and read retention beside benign task completion,
since lower attack rates matter only if the agent completes
legitimate tasks.

The three agents retain 86-91\% of consequences (38--40 of 44) while
completing 76-83\% of benign task attempts. The choice of agent moves
retention by at most 4.5 percentage points (two cases), no agent
neutralizes a class, and each loses at most 2 of the 40 realizable cases
(Table~\ref{tab:app-agents}).
Across the tested requests, all three largely carry the attacks through
to backend execution. Claude Code and Codex lose no realizable case to
refusal, input normalization, or execution failure. Each leaves 10\% of
its retained cases transformed, including changes to a non-essential
outer file name on a package-import attack. Grok Build loses 5\% of the
realizable cases to malformed or abandoned calls in multi-step
resource-handling workflows and leaves 8\% of its retained cases
transformed. Similar execution errors occur in its benign runs, whose
completion rate is 76\%. Doubling the caps recovers one of its two lost
cases (Appendix~\ref{app:controls}), showing budget sensitivity in
one observed loss.
Its lower retention accompanies lower benign completion, with no
case-level losses attributed to selective refusal. Claude Code and
Codex tie on retention and differ by one percentage point in completion.
Because the underlying models are opaque and the losses are few, these
observations do not establish a specific internal cause or a reliable
security ranking.

\begin{finding}{3}
Execution failures account for the observed case-level differences in
exploitability across agents. Similar failures also disrupt legitimate
tasks, linking lower attack success to reduced task reliability without
observed gains from selective rejection of malicious requests.
\end{finding}

\subsection{Impact of the Interface (RQ3)}
\label{sec:eval-rq3}

We then hold Claude Code fixed and vary only the capability design
(Figure~\ref{fig:sweeps}b,c), reading retention beside benign coverage
and task completion. The query-oriented design uses a fixed planner and
snapshot index, and any contrast holds for our implementations rather
than for the standards themselves.

The endpoint-oriented design retains 95\% of consequences (42/44) at
92\% benign coverage (46/50), the workflow-oriented reference 91\%
(40/44) at 84\% (42/50), and the query-oriented design 39\% (17/44) at
46\% (23/50); task completion is 89\%, 82\%, and 41\%, respectively.
The design thus moves retention by 56.8 percentage points, against at
most 4.5 across agents, but narrowing the surface removes exploit paths
and useful operations together. The one-factor sweeps do not measure
agent-by-design interactions, and another agent may behave differently
under the query-oriented design.

\begin{finding}{4}
Interface design changes exploitability more than agent choice in the
tested configurations. The largest reductions in attack exposure
accompany substantial losses of legitimate functionality, making
functional coverage essential to interpreting apparent security
improvements of an interface.
\end{finding}

Most of the difference occurs at the interface stage. The
endpoint-oriented design makes 98\% of cases expressible with almost no
subsequent loss. For the query-oriented design, 24 of its 27 losses
(55\% of all cases) occur at the interface stage, where no intent
reaches the required operation or slot filling fails to transmit the
required argument; the other 3 are C1-realizable cases that Claude Code
fails to exploit. To control for workflow coverage, we examine the 21
workflows (42\%) that all three designs support, fixed before any attack
result, which correspond to 17 cases. The ordering holds: the designs
retain 16, 15, and 9 cases (94\%, 88\%, and 53\%). The two
workflow-oriented losses here have a supported workflow but an
unexposed operation or argument. Of the 8 query-oriented losses, 6
reach the relevant intent but lose the attack argument in slot filling,
and 2 are Claude Code failures. The gap therefore reflects argument
delivery through typed slots rather than natural-language access as
such, conditional on our slot schema, planner, and request set.

\begin{finding}{5}
Supporting the same legitimate workflows does not imply delivering the
same attack inputs to the backend. Argument constraints and
query-planner behavior can prevent tested attacks from reaching their
effect even when the legitimate workflows remain available.
\end{finding}

\section{Discussion}
\label{sec:discussion}
\subsection{Implications for Agent-Facing Web Services}
\label{sec:discussion-services}

\textbf{Preserving Legitimate Functionality.}
As agents become a primary entry point to Web services, interface
restrictions determine which tasks users can delegate.
Our results show that lower exploitability can accompany the removal
of legitimate workflows.
The comparison on shared workflows shows that an operation can remain
available while attack inputs fail to reach the backend. In the
query-oriented design, these losses reflect both slot constraints and
planner execution. This distinction motivates interface designs that preserve required
business functions and constrain inputs according to their intended
use.
Agent safeguards require a similar distinction between rejecting
harmful actions and failing to execute tasks.
The execution failures observed in RQ2 affected both attacks and
legitimate tasks, providing little basis for treating lower attack
success alone as stronger protection.

\looseness=-1
\textbf{Authorization Granularity.}
Exposing a capability for a legitimate business task leaves open
which users may apply it to which objects and under what conditions.
A capability for retrieving orders, for example, still requires
checks on the caller's access to the requested order.
Binding calls to a user identity establishes who acts; object-level
authorization and business rules determine what that user may do.
Agent-facing interfaces can support these checks by binding
identity-dependent arguments and limiting unnecessary input control.
The backend must enforce the corresponding policy on each request,
including requests expressed through valid business arguments.
As services move beyond website navigation, these rules must remain
effective across every interface exposing the operation.

\looseness=-1
\textbf{Agent-Generated Call Sequences.}
The single step-completion case shows that an agent can supply prerequisite
calls omitted from a user's request and thereby complete an attack.
The security implications of delegation thus extend to the sequence
the agent constructs during execution.
Each call may invoke an ordinary business operation, while the
resulting sequence violates a rule governing eligibility, ordering,
or repeated execution.
This motivates enforcing workflow rules against the current
application state throughout execution.
Checks on the initial instruction or the availability of individual
tools alone cannot establish whether the completed workflow is
authorized.

\subsection{Implications for Security Evaluation}
\label{sec:discussion-evaluation}

Evaluating an agent-facing service requires explaining why an attack
fails and what legitimate functionality remains available.
An unavailable operation, an unsuccessful agent run, and an enforced
authorization check have different implications for service design.
Aggregate attack success can obscure these differences.
The comparisons in this study show the value of testing what an
interface permits independently of agent execution and verifying
the resulting effects on the application.
Reading these outcomes alongside benign coverage and completion
helps distinguish reduced exposure through functionality removal
from restrictions that preserve the intended task.

Future evaluations should also examine the call sequences agents
generate and repeat measurements as models and interfaces evolve.
An agent update may recover an execution failure that previously
prevented an attack, while an interface update may expose a missing
operation or argument.
Broader comparisons across agent--interface combinations would
clarify whether the effects observed in our separate sweeps persist
when both layers change.

\subsection{Threats to Validity}
\label{sec:discussion-validity}

\textbf{Representativeness.}
The benchmark covers 44 documented vulnerabilities across twelve
application families, with cases drawn from production applications
and deliberately vulnerable environments.
It supports controlled comparisons of known defects, but does not
estimate their prevalence across deployed Web services.

\textbf{Interface Construction.}
Retention depends on how much an interface exposes, and we built the
interfaces ourselves, so the 91\% reference figure is not an estimate
for deployed services. We derive them blind from benign workflows,
freeze them before any security test
(Section~\ref{sec:vela-interfaces}), report benign coverage beside
retention, and give a range across three designs (39\% to 95\%) rather
than a single point. The query-oriented results additionally depend on
our planner and slot constraints, and other implementations may expose
different operations or preserve different attack inputs.

\textbf{Execution and Measurement.}
Observed exploitability depends on the tested requests, agent versions,
and execution budgets. A security-oriented system prompt or approval by
a party other than the user may change agent-layer outcomes and is not
tested.
Case-level retention records success in at least one repetition and
does not imply consistent success across attempts.
Failures remain specific to the tested conditions, as illustrated by
the case recovered with a larger budget.
Independent oracles verify predefined security consequences; they
do not enumerate every possible harm.
Our one-factor comparisons identify differences around the reference
configuration, while interactions between agents and interface
designs remain to be evaluated.
The results apply to capability-based access under the stated threat
model; browser-driven access and attacks that compromise the agent
require separate investigation.

\balance
\section{Related Work}
\label{sec:related}
\textbf{Agents as Web Attackers.}
A growing body of work measures whether LLM agents can exploit Web
vulnerabilities on their own. Benchmarks such as
WebExploitBench~\cite{webexploitbench}, CVE-Bench~\cite{cvebench}, and
PACEbench~\cite{pacebench} deploy vulnerable applications as isolated
targets, and studies show that agents can hack live websites and exploit
disclosed one-day vulnerabilities from their
descriptions~\cite{fang_websites,fang_oneday}. Related systems configure
models for penetration testing and exploit
construction~\cite{pentestgpt,happe}. These studies examine agents'
vulnerability-discovery and exploitation capabilities under different
levels of prior knowledge and access. We reuse such environments as a source of validated
cases but study a different setting, in which a human attacker who already
knows the attack reaches the same backend through a default-configured
agent and an agent-facing interface. The question is not whether the agent
can find or build an exploit but whether this mediation carries a known
one through.

\textbf{Attacks on LLM Agents.}
A parallel line treats the agent itself as the target. Direct and indirect
prompt injection can override an LLM-integrated
application~\cite{greshake,promptinjection_liu}, benchmarks measure how
often tool-using agents follow injected
instructions~\cite{injecagent,agentdojo}, and emulated sandboxes expose
unsafe agent behavior at scale~\cite{toolemu}. Agent-in-the-middle
attacks intercept messages between cooperating
agents~\cite{he2025aitm}, whereas our agent in the middle sits between
a user and a backend. AgentHarm evaluates
harmful-task execution~\cite{andriushchenko2025agentharm}, while
ST-WebAgentBench evaluates safety and policy adherence during business
tasks~\cite{levy2026st}. BrowserART studies whether refusal training
transfers to browser agents~\cite{kumar2024refusal}, and the Cybersecurity
Refusal Framework examines refusals in cyber tasks in relation to
operational context~\cite{jones2026new}. Our study complements these
evaluations by comparing access paths to the same vulnerable backend,
separating interface effects from the external agent's execution, and
measuring the accompanying loss of legitimate functionality. We submit
requests that preserve the attack objective and exploit-relevant inputs
to a task assistant and verify the resulting backend consequences with independent oracles.

\looseness=-1
\textbf{Agent-Facing Web Access.}
Agents increasingly act over real Web interfaces, from grounded interaction
environments~\cite{webshop,webarena} to generalist agents on live
sites~\cite{mind2web}, built on reasoning-and-acting and tool-use
methods~\cite{react,toolformer}. Access schemes such as OpenAPI
descriptions, MCP~\cite{mcp}, and NLWeb~\cite{nlweb}, together with
commerce protocols such as UCP~\cite{ucp}, define how services expose
functionality to these agents. Security analyses of MCP catalog threats
such as malicious tool descriptions~\cite{hou2025mcp}, and prompt-to-SQL
injection shows that an LLM-integrated Web application can relay
attacker input into database queries~\cite{pedro2025p2sql}. This
literature describes how agent-mediated access works and the risks it
adds, whereas we measure its security consequence for existing
vulnerabilities through matched access paths and oracle-verified effects.

\section{Conclusion}
\label{sec:conclusion}
We presented \tool{}, a benchmark that compares native access, direct
capability invocation, and agent-mediated execution on the same
vulnerable backends under independent oracles, separating the effect of
the agent-facing interface from that of the agent. Across 44 cases, the
agent removed none of the attacks the reference interface let through,
and 91\% of the vulnerabilities retained their native consequence. The
capability design shifted retention far more than the choice of agent,
from 95\% under an endpoint-oriented design to 39\% under a
query-oriented one, though the latter also lowered benign coverage to
46\%. As agent-facing interfaces become a common path to the Web, what
each capability exposes, rather than the agent, is the main lever on
exposure in our implementation, with the admissible argument domain of
each capability as the most promising refinement to test.





\section*{Ethical Considerations}
\label{sec:ethics}
We exercise vulnerabilities only in applications that we deploy and
control, inside isolated environments with synthetic accounts and synthetic
data, and we test no live services and no third parties. The work involves no
human subjects or personal data, and the vulnerabilities come from existing
public benchmarks and disclosed advisories, so we report no new
vulnerabilities and make no disclosure claims. Our findings concern how
agents and interfaces handle known exploits rather than a flaw in a
specific product. A characterization of which
attacks survive agent mediation could inform misuse, but we judge its
defensive value to outweigh this risk, since it helps interface designers close exposed paths. 

\bibliographystyle{ACM-Reference-Format}
\bibliography{sample-base}

\appendix
\setlength{\intextsep}{8pt plus 2pt minus 2pt}
\setlength{\textfloatsep}{12pt plus 2pt minus 2pt}
\setlength{\floatsep}{8pt plus 2pt minus 2pt}
\section{Benchmark Composition}
\label{app:composition}

\begin{figure}[h]
    \centering
    \includegraphics[width=0.95\columnwidth]{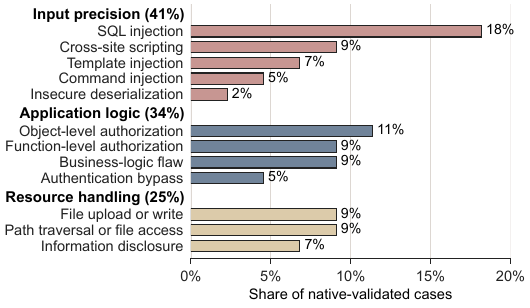}
    \caption{Share of native-validated cases by vulnerability category, grouped and coloured by analysis group.}
    \Description{Horizontal bars show 18 input-precision cases in five categories, 15 application-logic cases in four categories, and 11 resource-handling cases in three categories.}
    \label{fig:app-categories}
\end{figure}

\looseness=-1
Table~\ref{tab:app-roster} in Section~\ref{sec:eval-settings} summarizes
the application families and case provenance.
Figure~\ref{fig:app-categories} shows the
vulnerability categories and their analysis groups. A training challenge
is not an independent CVE and is admitted only after successful native
reproduction.

Representative identifiers include Dolibarr CVE-2024-5314 and
CVE-2024-5315, SuiteCRM CVE-2024-36412, Nextcloud CVE-2026-45281, Cacti
CVE-2024-25641 and CVE-2024-34340, Zabbix CVE-2024-22120, PrestaShop
CVE-2024-34716, and Moodle MSA-26-0044. The WordPress family draws on
eleven plugin CVEs, including the Web Directory Free SQL injection
CVE-2024-3552 and the Hash Form file-upload RCE CVE-2024-5084, each of
which needs its own workflow and interface mapping.

\section{Additional Results}
\label{app:results}

\begin{figure}[h]
    \centering
    \includegraphics[width=0.93\columnwidth]{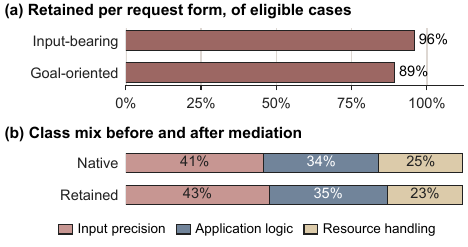}
    \caption{Reference configuration. (a) Share of eligible cases retained per request form. (b) Analysis-group shares among native and retained cases.}
    \Description{Input-bearing requests retain 96 percent of eligible cases and goal-oriented requests 89 percent. Group shares move from 41, 34, and 25 percent among native cases to 43, 35, and 23 percent among retained cases.}
    \label{fig:app-rq1}
\end{figure}

Figure~\ref{fig:app-rq1} complements RQ1 under the reference
configuration. Input-bearing requests succeed in 96\% of their
eligible cases and goal-oriented ones in 89\%; these use different
eligible sets. On the cases eligible for both forms, case-level
retention is identical. Mediation also leaves the class mix nearly
unchanged. Resource-handling cases fall from 25\% of the native set to
23\% of the retained set, consistent with the interface-level losses
reported in Section~\ref{sec:eval-rq1}.

Table~\ref{tab:app-agents} breaks down the agent sweep of RQ2 by outcome
kind. No case-level loss is attributed to refusal or input normalization,
and the only agent-layer losses are Grok Build's malformed or abandoned calls in
multi-step resource-handling workflows.

\begin{table}[h]
    \centering
    \caption{Agent-layer outcomes on 40 C1-realizable cases under the workflow-oriented design.}
    \label{tab:app-agents}
    \small
    \begin{tabular}{@{}lccc@{}}
        \toprule
         & Claude Code & Codex & Grok Build \\
        \midrule
        Transformed cases            & 4 & 4 & 3 \\
        Lost at the agent            & 0 & 0 & 2 \\
        \quad Refusal                & 0 & 0 & 0 \\
        \quad Input normalization    & 0 & 0 & 0 \\
        \quad Malformed or abandoned & 0 & 0 & 2 \\
        \bottomrule
    \end{tabular}
\end{table}

\balance
\section{Mechanism Controls and Sensitivity}
\label{app:controls}

Each control isolates one link among the interface, the agent, and the
request, and states what its result implies for attribution.

\textbf{Argument Replay.}
For each case realized by C2, we replay the agent-generated arguments
of the calls specified by the attacker, in order, from the same verified
snapshot through a scripted caller. This control tests whether the
specified calls suffice without the agent's additional steps. Under the
reference configuration, replay reproduces the effect in all realizable
cases except the crAPI step-completion case. Its C2 sequence comprises
an order lookup, return initiation, and return submission. Replaying
only the requested submission leaves the balance unchanged, whereas
the complete sequence credits it. The primary C1 baseline includes the
prerequisite calls and also succeeds. This contrast identifies the
dependence on the calls supplied by the agent.

\textbf{Input-Bearing and Goal-Oriented Requests.}
We compare requests that preserve the attack objective and relevant
inputs on the cases eligible for both forms. The two forms yield
identical case-level retention on this shared subset.

\textbf{Request Phrasing.}
We vary ordinary wording while keeping the attack objective and
exploit-relevant inputs fixed. Across 3 paraphrases per case, the outcome
is unchanged in 95\% of cases, indicating stability under these tested
wording changes.

\textbf{Budget Sweep.}
We vary the caps on calls, tokens, and time. Doubling each cap recovers
no additional case under the reference configuration and one under Grok
Build. One of Grok Build's two case-level losses is therefore sensitive
to the execution budget.

\textbf{Direct-Caller Baseline.}
For the step-completion case, the successful sequence contains 3 calls,
2 of which the attacker did not specify. A caller restricted to the
requested submission omits these prerequisites, whereas the complete C1
sequence includes them and succeeds. This control measures step
completion through the additional calls supplied by the agent.

\textbf{Interface-Comparable Subset.}
We compare agents on the C1-realizable cases under the fixed
workflow-oriented interface. Retention is 100\% under the reference
configuration, and the cross-agent spread stays within 5 percentage
points. These differences arise during agent execution on a shared set
of cases already realized by direct invocation.

\section{Paired Conditions and Case Mapping}
\label{app:paired}

\looseness=-1
Table~\ref{tab:app-paired} gives the paired C1/C2 outcome of every
configuration over its C1-realizable subset. These paired tables
exclude interface-stage losses, including structural restrictions and
query-planner execution losses. No case in the tested configurations
succeeds under C2 without C1. 

\begin{table}[h]
    \centering
    \caption{Paired C1/C2 outcomes on each configuration's C1-realizable subset ($n$), excluding interface-stage losses. \checkmark{} and $\times$ denote verified and unverified outcomes; $p$ is the exact two-sided McNemar $p$-value.}
    \label{tab:app-paired}
    \small
    \setlength{\tabcolsep}{3pt}
    \begin{tabular}{@{}lccccc@{}}
        \toprule
        Configuration & $n$ & C1\checkmark C2\checkmark & C1\checkmark C2$\times$ & C1$\times$ C2$\times$ & $p$ \\
        \midrule
        Claude Code, workflow & 40 & 40 & 0 & 0 & 1.0 \\
        Codex, workflow       & 40 & 40 & 0 & 0 & 1.0 \\
        Grok Build, workflow  & 40 & 38 & 2 & 0 & 0.50 \\
        Claude Code, endpoint & 43 & 42 & 1 & 0 & 1.0 \\
        Claude Code, query    & 20 & 17 & 3 & 0 & 0.25 \\
        \bottomrule
    \end{tabular}
\end{table}

\section{Oracle and Case Records}
\label{app:oracles}

For a confidentiality violation, the oracle marks the protected status of
synthetic data and checks that the attacker-visible channel returned it.
Presence of the value in the backend alone is not enough. For an
integrity violation, it compares persistent state before and after the
attempt. Other effects use trusted observations such as file state or
application event records. Each case record stores the native
reproduction, negative controls, the oracle predicate, actor and
ownership assertions, reset receipts, and per-condition traces.

Each executed attempt is classified as a verified success, a valid
failure, or an invalid run. Structural inexpressibility is recorded at
the case-mapping stage. Failed resets,
incompatible configurations, and inconclusive observations are invalid
and are reported separately. An invalid attempt is rerun from a fresh
snapshot up to 2 times, and in the reference configuration
2.7\% of the C2 attempts were invalid, all of which
were valid on the first rerun. An invalid run never counts as an
exploitation failure.

\end{document}